\documentstyle[11pt,epsfig]{article}  
\begin{document} 

%%%%%%%%%%%%%%%%%%%%%%%% NEW DEFINITIONS 
\def\la{\mathrel{\mathpalette\fun <}}
\def\ga{\mathrel{\mathpalette\fun >}}
\def\fun#1#2{\lower3.6pt\vbox{\baselineskip0pt\lineskip.9pt
\ialign{$\mathsurround=0pt#1\hfil##\hfil$\crcr#2\crcr\sim\crcr}}}  
\def\lrang#1{\left\langle#1\right\rangle}
%%%%%%%%%%%%%%%%%%%%%%%% END OF NEW DEFINITIONS 

\begin{titlepage} 

\begin{center}
{\Large \bf Rapidity distribution of jet quenching: effects on global 
observables\footnote{Contribution to the CERN Yellow Report on Hard Probes in 
Heavy Ion Collisions at the LHC.}}   

\vspace{5mm}

{\it I.P.~Lokhtin$^a$, S.V.~Shmatov$^b$ and P.I.~Zarubin$^b$} 

\vspace{2.5mm} 
    
$^a$ Institute of Nuclear Physics, Moscow State University, Russia \\ 
$^b$ Joint Institute of Nuclear Research, Dubna, Russia

\vspace{5mm} 

\end{center} 

\begin{abstract} 

We discuss modification in the rapidity distribution of the transverse
energy flow as a signal of medium-induced partonic energy loss in
ultrarelativistic heavy ion collisions. A scan of the global energy flow and 
jet rates in the wide rapidity region might provide the important 
information about the pseudo-rapidity size of a dense medium created in heavy
ion collisions at the LHC. 

\end{abstract}

\end{titlepage}

Among other effects originated by medium-induced parton energy loss one may expect a 
observable modification in the rapidity distributions of the transverse energy flow and 
charged multiplicity, $dE_T/d \eta$, $dE_T^{\gamma}/d \eta$, and 
$dn_{ch}/d \eta$~\cite{Savina:1999,Damgov:2001}. 

\begin{figure}[htb]
\begin{center} 
\resizebox{150mm}{150mm} 
{\includegraphics{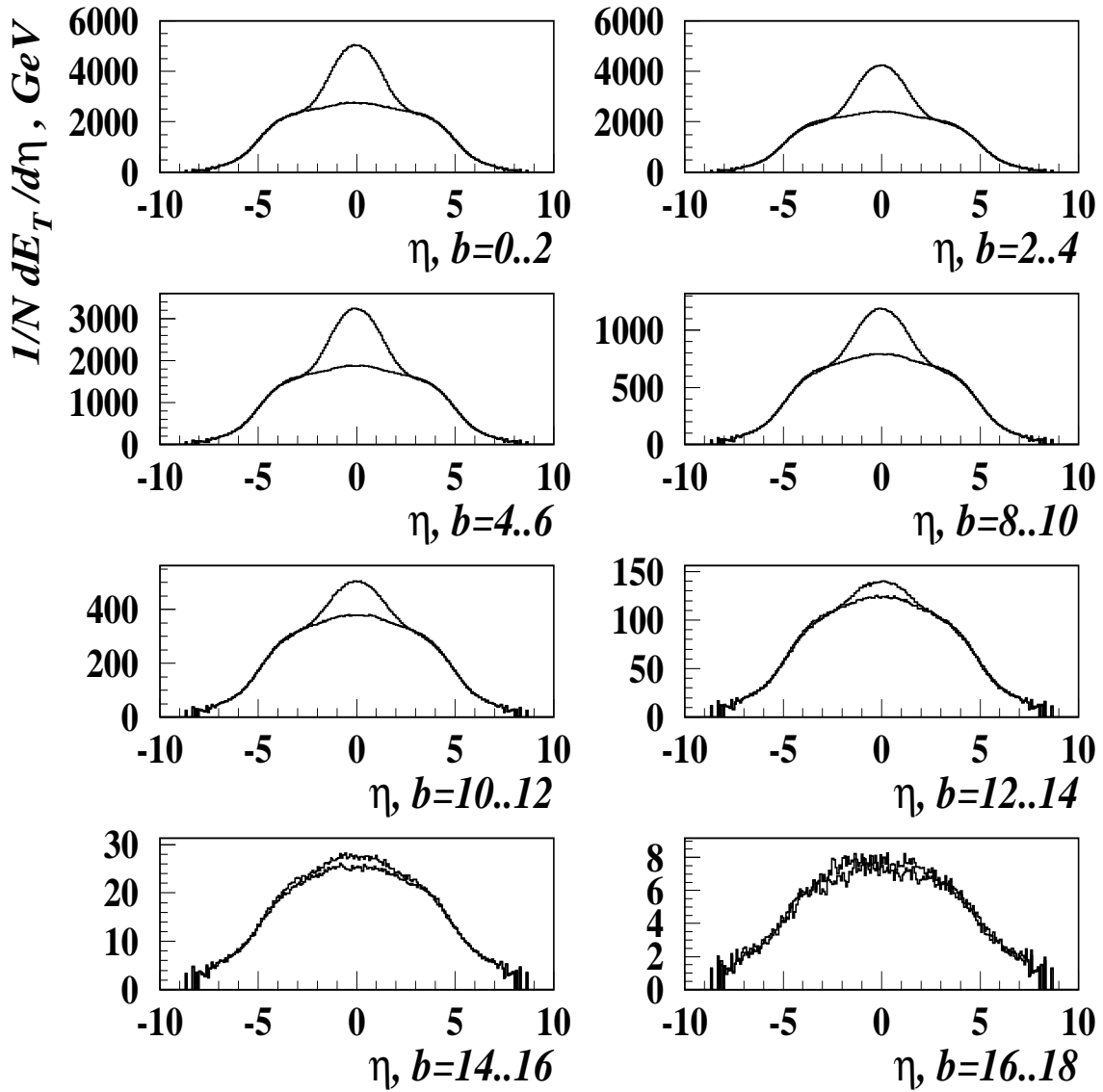}} 
\caption{\small Normalized differential distribution of the total transverse energy 
$dE_T/d\eta$ over pseudo-rapidity $\eta$ for $10.000$ minimum bias Pb$-$Pb collisions 
at $\sqrt{s}= 5.5 A$ TeV for various impact parameters. The two 
cases are included: with jet quenching (top histogram) and without jet quenching 
(lower histogram).} 
\label{gl:fig1}
\end{center} 
\end{figure}

Indeed, an indication is found for ultrarelativistic energy domain concerning the 
appearance of a wide bump in the interval $-2 \la \eta \la 2$ over a pseudo-rapidity 
plateau of such distributions due to jet quenching. Figure~\ref{gl:fig1}
demonstrates the evolution of the effect with Pb$-$Pb collision 
impact parameter variation (HIJING~\cite{hijing} prediction for the LHC energy, 
$\sqrt{s}= 5.5 A$ TeV). One can see that even peripheral 
Pb$-$Pb collisions show the effects of energy loss with the central enhancement still 
evident at impact parameters up to $12$ fm. Since jet quenching due to final state 
re-interactions is effective only for the mid-rapidity region (where the initial energy 
density of minijet plasma is high enough), very forward rapidity region, $|\eta| 
\ga 3$, remains practically unchanged. A scan of collisions of different nuclear
systems provides an additional test of jet quenching. Because smaller nuclei require a
shorter transverse distance for the partons to traverse before escaping the system,
the central enhancement due to energy loss decreases with system size as obvious from
the comparison with and without energy loss. Note that although the effect has only 
been shown here for the global $E_T$ distributions $dE_T/d \eta$, qualitatively the 
same picture is seen when neutral or charged particle production is studied instead of
$E_T$.   

The greater the medium-induced energy loss, the more transverse energy is piled up at
central $\eta$ values, leading to an increase in energy density or "stopping" in the
mid-rapidity region, in contradiction to the assumption of nuclear transparency. 
It is interesting to note that the qualitatively same results shown in
figure~\ref{gl:fig1} can be obtained using VENUS generator~\cite{venus} and Parton
Cascade Model VNI~\cite{vni}, in spite of the physics of the VENUS nucleon rescattering 
or VNI parton rescattering modes is very different than that of the radiative energy 
loss mechanism in HIJING. This may be due to the fact that the many kinds of nuclear 
collective effects like final state re-interactions are effective forms of "nuclear 
stopping". 

One can consider rapidity spectrum of jets or high-p$_T$ products of jet
fragmentation (the latter case has been discussed for RHIC energy in a 
paper~\cite{Polleri:2001}). The pattern of rapidity distribution of only hard jets (say, 
at $E_T^{jet}>100$ GeV) is expected to be rather opposite to
that for global $dE_T/d \eta$ spectrum. Since jet quenching due to in-medium parton 
energy loss (more exactly, the part of energy loss which is outside the definite jet
cone size, see for discussion~\cite{Lokhtin:1998,Baier:1998}) is strongest in 
mid-rapidity, the maximal suppression of jet rates as compared to what is expected from 
independent nucleon-nucleon interactions extrapolation can be observed at central 
rapidity, while the very forward rapidity domain remains again almost unchanged. 
Thus analyzing the correlation between rapidity distributions of global energy 
(particle) flow and hard jets, by scanning the wide rapidity region 
(up to $|\eta|\le 5$ under acceptance of CMS experiment~\cite{Baur:2000}), might 
provide the important information about the pseudo-rapidity size of dense QCD-matter 
area. 

It is pleasure to thank M.~Bedjidian, D.~Denegri, L.I.~Sarycheva and U.~Wiedemann for 
encouraging and interest to the work. Discussions with J.~Damgov, V.~Genchev, 
S.~Kunori, M.V.~Savina, N.V.~Slavin and A.M.~Snigirev are gratefully acknowledged.

\vskip 1 cm 

\end{document}